\documentclass[journal=langmuir,manuscript=article]{achemso}

\usepackage[version=3]{mhchem} 
\usepackage{comment}
\usepackage{siunitx}

\newcommand{\et}{\textit{et al.}}

\author{Lijun Thayyil Raju}
\affiliation{Physics of Fluids Group, Faculty of Science and Technology, University of Twente, 7500 AE Enschede, The Netherlands}

\author{Christian Diddens}
\affiliation{Physics of Fluids Group, Faculty of Science and Technology, University of Twente, 7500 AE Enschede, The Netherlands}

\author{Yaxing Li}
\affiliation{Institute of Fluid Dynamics, Department of Mechanical and Process Engineering, ETH Z\"urich, 8092 Z\"urich, Switzerland}

\author{Alvaro Marin}
\affiliation{Physics of Fluids Group, Faculty of Science and Technology, University of Twente, 7500 AE Enschede, The Netherlands}

\author{Marjolein N. van der Linden}
\affiliation{Canon Production Printing Netherlands B.V., 5900 MA Venlo, The Netherlands}
\alsoaffiliation{Physics of Fluids Group, Faculty of Science and Technology, University of Twente, 7500 AE Enschede, The Netherlands}

\author{Xuehua Zhang}
\affiliation{Department of Chemical and Materials Engineering, University of Alberta, Edmonton, Alberta T6G 1H9, Canada}
\alsoaffiliation{Physics of Fluids Group, Faculty of Science and Technology, University of Twente, 7500 AE Enschede, The Netherlands}

\author{Detlef Lohse}
\affiliation{Physics of Fluids Group, Faculty of Science and Technology, University of Twente, 7500 AE Enschede, The Netherlands}
\alsoaffiliation{Max Planck Institute for Dynamics and Self-Organisation, 37077 G\"ottingen, Germany}
\email{d.lohse@utwente.nl}

\title[An \textsf{achemso} demo]
{  Evaporation of a sessile colloidal water-glycerol droplet: Marangoni ring formation
}

\abbreviations{empty}

\begin{document}

\begin{tocentry}

\includegraphics{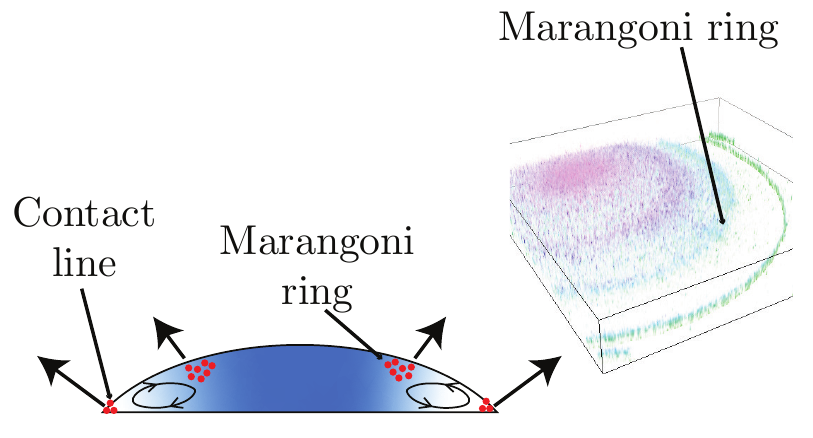}

\end{tocentry}

\begin{abstract}
The transport and aggregation of particles in suspensions is an important process in many physicochemical and industrial processes.
In this work, we study the transport of particles in an evaporating binary droplet. Surprisingly, the accumulation of particles occurs not only at the contact line (due to the coffee-stain effect) or at the solid substrate (due to sedimentation), but also at a particular radial position near the liquid-air interface, forming a ``ring'', which we term as the \textsl{Marangoni ring}.
The formation of this ring is primarily attributed to the solutal Marangoni flow triggered by the evaporation dynamics of the water-glycerol droplet. Experiments and simulations show fair agreement in the volume evolution and the general structure of the solutal Marangoni flow, that is, the \textsl{Marangoni vortex}. Experiments show that the location of the Marangoni ring is strongly correlated with the Marangoni vortex. However, finite element numerical simulations fail to describe the particle distribution seen in the experiments.
Interestingly, the particles not only accumulate to form the Marangoni ring but also assemble as colloidal crystals close to the liquid-air interface, yielding iridescence. The formation of the colloidal crystals in the experiments is strong evidence that non-hydrodynamic interactions, which are not represented in the simulations, also play a significant role in our system.

\end{abstract}


\section{Introduction}
Multicomponent sessile droplets containing colloidal particles are ubiquitous in nature and technology. Even though a colloidal droplet seems to be a simple system, evaporation leads to a complex physicochemical scenario \cite{Lohse2020,Sefiane2022} involving evaporation-driven flows \cite{Deegan1997}, segregation \cite{Li2018,Kim2018}, phase-separation \cite{Tan2016,ThayyilRaju2021a,Guo2021,ColinetDemixing2020}, and flow patterns driven by gradients in concentration \cite{Christy2011,Sempels2013a,Marin2019}, temperature \cite{Hu2006}, and density \cite{Li2019c}.
The fluid flow inside the droplet strongly determines the transport of the dispersed particles. Understanding the transport of particles is important in scenarios such as biofluid droplets \cite{Guo2021,Carola_2022}, electronics \cite{Eom2014,Sirringhaus2000,Vanmaekelbergh2015}, inkjet printing \cite{Lim2008,Hu2020,Du2020,Lohse_2022_fip}, and catalysis \cite{Hou2020}. 

The radially oriented capillary flow in a pinned evaporating droplet can transport particles toward contact line, forming the so-called coffee-stain deposits \cite{Deegan1997}. 
However capillary flow is not the only evaporation-driven flow that can appear in a droplet. Preferential evaporation at the contact line of a multicomponent droplet can also lead to a gradient in the interfacial composition along the liquid-air interface, generating interfacial Marangoni shear in,  for example,  salt mixtures \cite{Marin2019} or surfactant \cite{Still2012,Sempels2013a} solutions. Thermal gradients can also easily appear, leading to interfacial thermal Marangoni shear stresses \cite{Hu2006,Ristenpart2007}.
These gradients in composition and/or temperature, lead to gradients in surface tension producing solutal and/or thermal Marangoni flow \cite{Lohse2020,Lohse_2022_fip,Sefiane2022}. 
These interfacial flows add up to complex flow patterns within the evaporating droplets. Occasionally, such flow patterns might prevent the particles to reach the contact line, suppressing the formation of the coffee ring\cite{Hu2006,Still2012}. In other instances, such flows might provide an alternate route for the particles to reach the contact line\cite{Marin2019,Bruning2020}.

However, Marangoni flow can also cause the accumulation of particles at the liquid-air interface. For example, thermal Marangoni flows are known to drag particles away from the contact line toward the droplet's apex \cite{Rossi2019}. Particles accumulate in the central region of the interface, forming a cap with a dense border, which we will term from now on as ``Marangoni ring''.
Such a cap with visible rings was already described by Deegan et al. \cite{Deegan2000}, and more recently studied by Parsa et al.\cite{Parsa2015} and Zhong et al.\cite{Zhong2016}, when particle laden droplets were evaporating on substrates at high temperatures. Rossi et al.\cite{Rossi2019} also found an inner ring in water droplets having small amounts of mineral salts. Particle laden droplets evaporating at very low pressures also show an inner ring \cite{Ren2020}. In all these cases, the thermal Marangoni flow transports the particles along the liquid-air interface, away from the contact line and toward the inner ring. Similarly, when the droplet contains surfactants \cite{Still2012,Sempels2013a} or surfactant-like polymers \cite{Seo2017}, the solutal Marangoni flow transports the particles along the liquid air interface and toward the inner ring. 

In this study we will focus on the emergence of Marangoni rings in droplets containing a mixture of two liquids. In such a binary droplet, the properties of the individual liquids, such as their volatility \cite{Parimalanathan2021}, mass density\cite{Li2019c}, surface tension \cite{Katre2021,Gurrala2021,Parimalanathan2021}, and viscosity \cite{Du2020} determines the nature of the fluid flow. Mixtures of water and glycerol are of special interest, for instance in inkjet printing, laboratory studies \cite{Sousa1995,Forney1992}, and biological processes \cite{Storey1997,Dashnau2006}. The high viscosity, miscibility in water, low cost, and non-toxic nature makes glycerol an excellent thickener for consumer products.

In this work, we show that a particle laden binary droplet of water and glycerol shows a very distinct Marangoni ring similar to droplets containing only one liquid. To the best of our knowledge, this is the first detailed study of the spatio-temporal evolution of the Marangoni ring in a binary droplet. We use high-resolution confocal microscopy and explore a wide range of initial glycerol concentrations and different particle types. Three-dimensional velocity measurements and finite element simulations are performed to understand the formation of the Marangoni ring in this system. 
Our experiments show colloidal crystallization and iridescence close to the Marangoni ring, a strong indication of the relevance of non-hydrodynamic interactions between the different particles and between the particles and the liquid-air interface.

\section{Experiments and Methods}

\subsection{Chemicals and Materials}
The following chemicals were used as received: Acetone (Boom BV, $\mathrm{>}$ 99.5\% (v/v), technical grade), Ethanol (Boom BV 100\%(v/v), technical grade), Glycerol (Sigma-Aldrich, $\geq$ 99.5\%, ACS reagent). Non-fluorescent silica particles (diameter = 0.970 \si{\micro \metre}, SD=0.029 \si{\micro \metre}, aqueous suspension), non-fluorescent polystyrene (PS) particles (diameter = $\mathrm{1.05 \pm 0.04}$ \si{\micro \metre}, aqueous suspension), and fluorescent PS particles (PS-FluoRed-Particles, diameter = 0.980 \si{\micro \metre}, SD = 0.04 \si{\micro \metre}, $\mathrm{abs/em = 530/607 nm}$) were obtained from Microparticles GmbH, Germany. Fluorescent Rhodamine B-labeled silica particles having diameter of 800 nm (diameter=$\mathrm{776 \pm 56}$ nm in TEM, $\zeta$-potential of $\mathrm{-56 \pm 1}$ mV, concentration: 23.5 mg/g of solution) were obtained from the Max Planck Institute of Polymer Research, Mainz, Germany. The details of the synthesis and characterization of the particles can be found in Thayyil Raju \et\cite{ThayyilRaju2021a}. All these concentrated particle dispersions were stored at a temperature of $\mathrm{4 ^\circ}$C before use. Milli-Q water was produced by a Reference A+ system (Merck Millipore) at 18.2 M$\Omega$ cm and 25$^{\circ}$C.
 
\textbf{Preparation of Dispersion of Water, Glycerol, and Particles.} 
To study the transport of particles in a binary droplet, we prepared a dispersion of particles, water, and glycerol. A small amount of the dispersion of particles was diluted in water and sonicated for 5 mins. Thereafter, the desired amount of glycerol was added into the diluted dispersion and sonicated again for 5 mins. The weight fraction of glycerol, $w_{g,i}$, in the dispersion was 0, 0.5, 5, 25, and 50\%. The concentration of the particles is kept constant at 0.1 wt \%.

\textbf{Preparation of the Substrate.} 
The glass substrates were carefully cleaned before the evaporation experiments. The substrates were first wiped with ethanol wetted tissue and water wetted tissue. Thereafter, the glass slides were sonicated in a bath of acetone for 5 min. This was followed by sonication in a water-bath for another 5 min. The slides were finally rinsed with ethanol and water, and blow-dried using nitrogen, before storing the glass for further experiments. Water droplets had a static contact angle of $\mathrm{10^{\circ}-20^{\circ}}$ on the cleaned glass substrates.

\subsection{Side-View and Top-View Visualization of the Evaporating Droplet}
To study the particle transport in the binary droplet, a 0.6 \si{\micro \litre} droplet of the particle laden water-glycerol dispersion was pipetted on the cleaned glass substrate. The droplet was illuminated using LED light sources. The droplet was viewed from the side and the top using two separate cameras (Figure S1a). For side-view, we used a monochrome 8-bit CCD camera (XIMEA, MQ013MG-ON, ~4 \si{\micro \metre}/pixel, 1 frame/s), attached to a Navitar 12$\times$ adjustable zoom lens. For top-view, a CMOS color camera (Nikon D750, 1920 $\times$ 1080 pixels, $\mathrm{\approx 4}$ \si{\micro \metre}/pixel, 24 frames/s) was used, attached with a Navitar 12$\times$ adjustable zoom lens. The ambient temperature and humidity were measured as $\mathrm{ 20 \pm 1 ^\circ}$C and $\mathrm{ 50 \pm 5 \%}$, respectively, using a thermo-hygrometer (OMEGA; HHUSD-RP1).

\textbf{Image Analysis}
The side-view images are processed using FIJI \cite{schindelin2012fiji} and in-house MATLAB codes, to determine the profile of the droplet. The volume of the droplet is determined by assuming that the droplet has a spherical-cap shape. Finally, the mass of water in the droplet is estimated using the volume of the droplet, initial composition of the water-glycerol mixture, and the composition-dependent density of the mixture, and by assuming that the amount of glycerol in the droplet is conserved.

\subsection{Measuring the Spatio-Temporal Particle-Distribution in the Droplet}
A laser scanning fluorescence confocal microscope (Nikon confocal microscopes A1 plus system) was used to obtain the spatio-temporal distribution of the particles inside the evaporating droplet (Figure S1b). A laser with a wavelength of 561 nm, filter cube (590/50 nm), and DU4 detector were used to observe the fluorescent Rhodamine B labeled silica particles. The confocal microscope was operated in resonant mode with a Plan Fluor 10$\times$ DIC objective and having in-plane resolution of 2.5  \si{\micro \metre}/pixel. The 3-dimensional particle distribution is reconstructed by combining images from horizontal planes which are separated by approximately 5 \si{\micro \metre} distance. It takes approximately 1 \si{\second} to collect the signals from all the horizontal planes for each time step, which is sufficiently fast as compared to the evaporation time of hundreds of \si{\second}. The confocal microscope was also operated separately in the Galvano mode at low magnification (with a Plan Fluor 4$\times$ objective having in-plane resolution of 6.2 \si{\micro \metre}/pixel) at 5 \si{\second} per frame, to obtain particle distribution in the entire droplet within the field of view. 

\subsection{Measuring the Flow-Field in the Droplet}
To determine the evaporation-induced flow in the droplet, we performed micro particle image velocimetry (\si{\micro PIV}) and three dimensional particle tracking velocimetry.

\textbf{\si{\micro PIV}.}
2D velocity fields are measured by  \si{\micro PIV}  using fluorescently labelled silica particles dispersed in the droplet and observed using the confocal microscope in resonant mode with a Nikon 20$\times$ water immersion objective, of numerical aperture 0.5, yielding a resolution of 1.24 \si{\micro \metre}/pixel at 15 and 30 frame/s. The velocity field was obtained in a squared region of 0.636 mm $\times$ 0.636 mm (512 pixel $\times$ 512 pixel) close to the contact line, in a horizontal plane approximately 10 \si{\micro \metre} above the solid substrate. The depth of correlation is estimated by the spread of the fluorescent microparticle's image along the optical axis,  which extends up to 10 \si{\micro \metre} approximately.
Cross-correlations and the velocity field were obtained using PIVLab 2.53 (running in MATLAB) \cite{thielicke2014pivlab,Thielicke2014, thielicke2021particle} (see more details in Supporting Information, Section S3). The velocity field was averaged temporally over 0.33 \si{\second} to obtain the final velocity distribution. To determine the average radial velocity as a function of the distance from the contact line, a rectangular strip of approximately 0.2 mm (along the contact line) $\times$ 0.6 mm (normal to the contact line) was selected, and the velocities were binned based on the distance from the contact line, with a bin size of 0.04 mm. Finally, the mean and standard deviation of the radial velocity is plotted.

\textbf{3D Particle Tracking Velocimetry.}
The three dimensional structure of the velocity field inside the droplet was measured by using General Defocusing Particle Tracking Velocimetry (GDPTV). To perform GDPTV, 1 \si{\micro \metre}-diameter fluorescent PS particles were dispersed in a mixture of water and glycerol (particle concentration of approximately $\mathrm{ 5 \times 10^{-5} \%}$ w/v). The motion of these particles inside an evaporating water-glycerol droplet was observed using an ECLIPSE Ti2 inverted microscope (Figure S1c). In GDPTV, the positions of the particles in the optical axis are obtained based on the characteristic particle image shapes at different distances from the focal plane. The main idea of the GDPT algorithm is to rely on a reference set of experimental particle images at known depth positions which is used to predict the depth position of measured particle images of similar shape. The recorded microparticle images are processed using DefocusTracker 2.0.0 (running in MATLAB) \cite{barnkob2021defocustracker,Rossi_2020}. 

\subsection{Finite Element Method (FEM) simulations of evaporating droplets}
Numerical simulations are performed using an axisymmetric finite element method (FEM) approach to further understand the evaporating particle-laden water-glycerol droplet. The assumption of asymmetry is justified for evaporating water-glycerol droplets as there are no instabilities that break the axisymmetry \cite{Diddens2017,Li2019c}. On the contrary, such axisymmetry-breaking instabilities occur in evaporation of water-ethanol droplets \cite{diddens_jfm_2017,Christy2011,bennacer_sefiane_2014} or condensation of water-vapor on a pure glycerol droplet\cite{Diddens2017,Stone_Glycerol2016}. To model the evaporation and the evaporation-induced flow within the droplet, the governing differential equations of continuity, mass transport, and momentum transport (Navier-Stokes) are solved for the water-glycerol mixture inside the droplet as well as for the water-vapor in the air surrounding the droplet. In the gas phase, only diffusive transport of water vapor is solved, i.e. Stefan flow and natural convection are disregarded. By using Raoult's law generalized by activity coefficients predicted by AIOMFAC\cite{Zuend2011}, the saturated vapor at the liquid-gas interface as function of the liquid composition can be imposed, whereas the ambient vapor concentration is imposed with a Robin boundary condition at the far field, mimicking an infinite domain \cite{Diddens2017}. Glycerol is assumed to be non-volatile due to its low vapor pressure.

The diffusive flux of water vapor at the interface is used to account for the volume loss and the local composition change in the droplet domain. The contact line is assumed to be pinned, where a tiny slip length (1 \si{\micro \metre}) at the substrate is used to resolve the incompatibility of a no-slip boundary condition at the substrate and the kinematic boundary condition with evaporation at the liquid-gas interface. In the droplet, the properties of the mixture, that is, mass density, dynamic viscosity and diffusivity, are known functions of the local fluid composition, which have been obtained by fitting experimental data \cite{Cheng2008,DErrico2004,Takamura2012}. The same applies for the surface tension at the liquid-gas interface, giving rise to solutal Marangoni flow. 

To monitor the distribution of particles, an advection-diffusion equation for a passive dilute continuous particle field is considered. The diffusivity of the particles is estimated using the Stokes-Einstein equation and at the liquid-gas interface, the local increase of the particle concentration due to water evaporation is implemented. Of course, this approach does not allow for accurate particle-particle or particle-fluid interactions and hence loses its validity once the particle concentration exceeds the dilute limit. As long as the particles are dilute, however, reasonable results can be expected.

Finally, to accurately track the interface motion, the FEM uses an arbitrary Lagrangian-Eulerian (ALE) approach, where the mesh nodes are co-moved with the interface motion. The strongly coupled set of equations is discretized in space by triangular Taylor-Hood elements with first order spaces for the vapor field, liquid composition and pressure and second order basis functions for the velocity. The equations are solved monolithically with a fully implicit backward differentiation formula of second order for the dynamic time stepping. The numerical method has been successfully applied in a widespread range of evaporating multicomponent droplets\cite{Li2018,Li2019c,Li2020,diddens_li_lohse_2021}. More details on the method can be found in these references. The implementation is based on the free finite element framework \textsl{oomph-lib} \cite{Heil2006}.

To assess potential influences on the flow in the gas phase (Stefan flow, natural convection) and thermal effects (evaporative cooling, thermal Marangoni flow), these mechanisms have also been considered in additional simulations. However, the consideration of these mechanisms did neither influence the flow inside the droplet nor the volume evolution in a relevant manner, so that these mechanisms have been disregarded for simplicity.

\section{Results and Discussions}

\subsection{Evaporation Dynamics of Particle-Laden Liquid Mixture Droplets}

\begin{figure}
    \centering
    \includegraphics[scale=0.89]{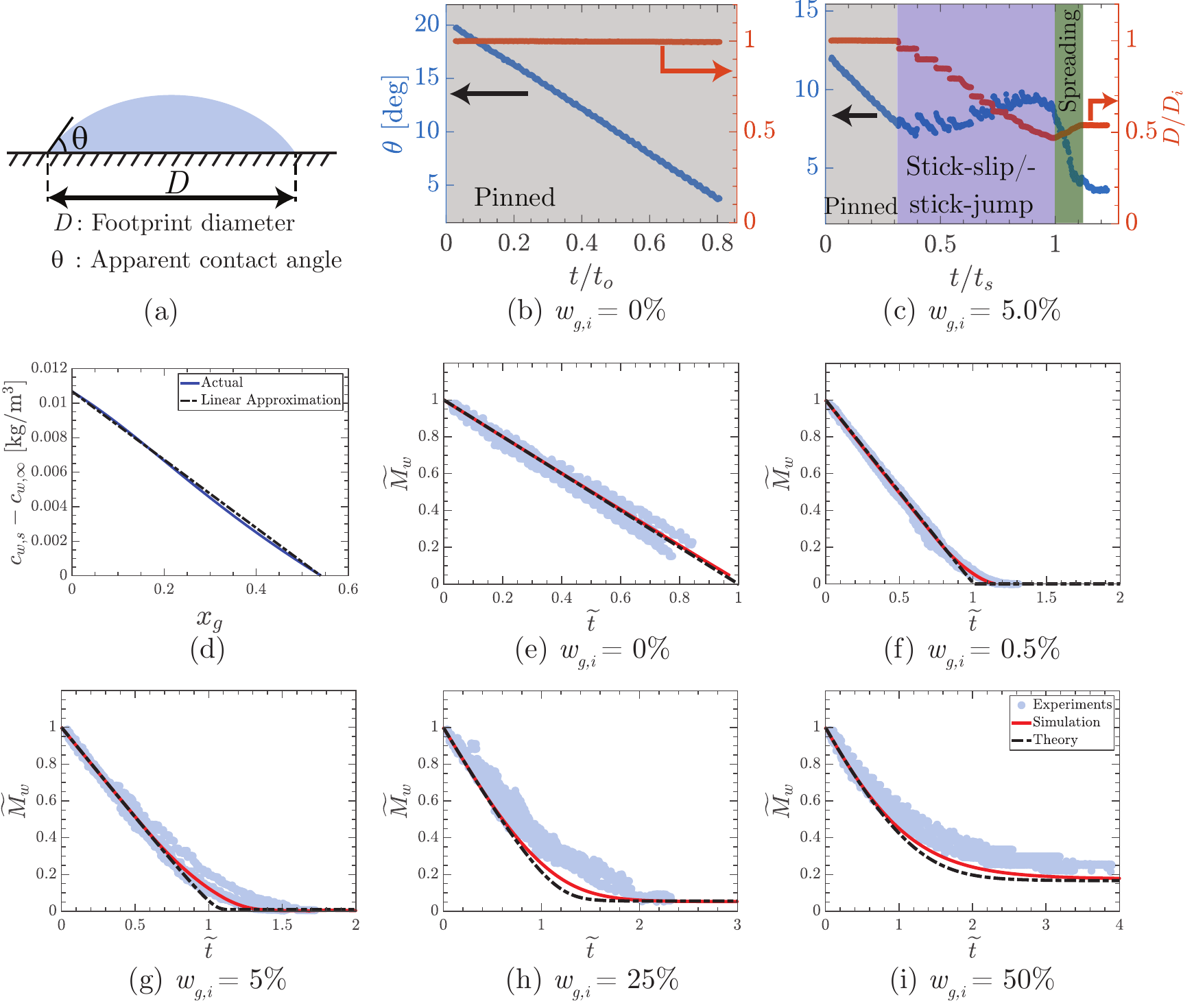}
    \caption{ \textbf{Evaporation dynamics of a particle laden water-glycerol droplet.}
    (a) Schematic of a sessile droplet showing the footprint diameter ($D$) and the apparent contact angle ($\mathrm{\theta}$) on the glass substrate. (b, c) Plots showing the variation of $D$ (normalized with initial diameter $D_i$) and $\mathrm{\theta}$ with normalized time for $w_{g,i}$ = 0\% and 5\%. (d) Plot of variation in \(c_{w,s}-c_{w,\infty}\) with glycerol mole fraction \(x_{g}\) - blue solid line (the actual variation without any approximation) and black dash-dotted line (linear approximation). (e-i) Plots showing the variation of the non-dimensional mass of water (\(\widetilde{M}_{w}\)) with non-dimensional time (\(\widetilde{t}\)) for different initial glycerol weight fractions - experiments (blue circles ), simulations (red solid lines), and analytical model (black dash-dotted lines). In experiments, the mass of water is estimated from the side-view imaging (see Experiments and Methods). All the droplets initially contained 0.1 wt\% of 800 nm silica particles.
    }
    \label{fig:evaporation_plots}
\end{figure}

The evaporation dynamics in liquid mixture droplets are far from trivial \cite{SAZHIN2017,Lohse2020,Sefiane2022}. Theoretical models, like  the celebrated model by Popov \cite{Popov2005,Gelderblom2011} are available to accurately predict the evaporation rate of sessile droplets of pure liquids. However, to the extent of our knowledge, there is no simple and generic analytical model available for evaporating sessile droplets of mixtures.
Therefore, to understand the formation of the Marangoni ring, we first study the evaporation dynamics of the droplet (Figure \ref{fig:evaporation_plots}).

The presence of contact line dynamics during the evaporation introduces an additional level of complexity to the process. 
Fortunately, for each of the compositions explored, the different modes of contact line motion are highly reproducible. Figure \ref{fig:evaporation_plots} a-c shows the contact line dynamics of a particle-laden water droplet and water-glycerol droplet ($w_{g,i}=$ 0 and 5\% respectively). In both cases, the diameter of the silica particles was 0.8 \si{\micro \metre}. For the particle-laden water droplet, the evaporation time $t$ is normalized by time $t_o$ ($ 340 \pm 50$ s), when evaporation is complete (Figure \ref{fig:evaporation_plots}b). For the particle-laden water-glycerol droplet, the evaporation time $t$ is normalized by the time $t_s$ ($ 370 \pm 50$ s), when the spreading of the droplet starts (Figure \ref{fig:evaporation_plots}c).

The contact line of a particle-laden water droplet remains pinned for most part of the evaporation, and depins only after $t/t_o = 0.90 \pm 0.03$ (Figure \ref{fig:evaporation_plots}b). On the contrary, the contact line of a particle-laden water-glycerol droplet ($w_{g,i} = 5 \%$) is pinned only during the initial period of evaporation (Figure \ref{fig:evaporation_plots}c and Supporting Information, Video V1). Thereafter, the contact line undergoes stick-slip and stick-jump motion \cite{Dietrich2015} (Figure \ref{fig:evaporation_plots}c). Close to the end of evaporation, the contact line spreads outward \cite{Parimalanathan2021} (Figure \ref{fig:evaporation_plots}c). Also for $w_{g,i}$ = 0.5 \% and 25 \%, the contact lines of the droplets go through the pinned, the stick-slip/stick-jump, and the spreading phases (See Supporting Information, Figure S2).

To further quantify the evaporation dynamics, we study the change in mass of water $M_w$ versus time $t$ (Figure \ref{fig:evaporation_plots} e-i). $M_w$ is normalized as $\widetilde{M_{w}}$ using the initial mass $M_{w,i}$ of water and time is normalized as $\widetilde{t}$ using a characteristic time scale, $\tau$. To determine $\tau$, we use the Popov's model \cite{Popov2005} of pure liquid droplet evaporation as below

\begin{equation}
  \frac{\mathrm{d}M_{w}}{\mathrm{d}t} = -\pi D_{v,a} R (c_{w,s}-c_{w,\infty}) f(\theta) \, ,  \label{eqn:popv}
\end{equation}
where \( D_{v,a}\) is the diffusion coefficient of water vapor in the air, \( R\) is the radius of the droplet, \( c_{w,\infty}\) is the vapor concentration of water far away from the droplet (corresponding to room humidity), \(c_{w,s}\) is the vapor concentration of water at the surface (liquid-air interface) of the droplet, \(\theta\) is the contact angle of the droplet, and \(f(\theta)\) is a known function of the contact-angle \cite{Popov2005}.

For a water-glycerol mixture, \(c_{w,s}\) is not a constant, but depends on the composition, in accordance with Raoult's law \cite{Raoult1887} and the non-ideal behavior of water-glycerol mixtures. Thus, \(c_{w,s} = x_{w} \psi_{w} c^{0}_{w,s}\), where \(x_{w}\) is the mole-fraction of water, \(\psi_{w}\) is the activity coefficient, and \(c^{0}_{w,s}\) is the saturation vapor concentration of pure water at the air-water interface. 

We approximate the dependence of \(\psi\) on \(x_{g}\) to be linear, such that 
\begin{equation}
    \psi (x_{g}) = 1 - A x_{g} \, . \label{eqn:si_psi_def}
\end{equation}

\noindent Further, by neglecting quadratic terms of \(x_{g}\), we get 
\begin{equation}
    c_{w,s}-c_{w,\infty} = (1-(1+A)x_{g})c^{0}_{w,s}-c_{w,\infty} \, . 
\end{equation}

\noindent We choose A such that \(c_{w,s}-c_{w,\infty}\) becomes zero at the same value of \(x_{g}\) as when using the exact value of \(\psi\) without linearization (Figure \ref{fig:evaporation_plots}d and Supporting Information, Figure S3). Hence,

\begin{equation}
    \frac{\mathrm{d}M_{w}}{\mathrm{d}t} = -\pi D_{v,a} R ((1-(1+A)x_{g})c^{0}_{w,s}-c_{w,\infty}) f(\theta) \, .  
\end{equation}

Because the contact angles are very low, we can take \(f(\theta) = 4/\pi\) \cite{Brutin2011}. Furthermore, the contact-line does not stay in constant contact angle or constant contact radius mode throughout evaporation, but rather shows various contact angle modes. Hence, for the purpose of solving this differential equation analytically, we take $R$ to be a constant (equal to the initial radius). We also assume that the composition in the droplet is spatially uniform. Thus,

\begin{equation}
    \frac{\mathrm{d}\widetilde{M}_{w}} {\mathrm{d}\widetilde{t}} = \frac{\alpha}{\beta + \gamma  \widetilde{M}_{w}} - 1 \, , \label{eqn:diff_en}
\end{equation}

\begin{equation}
    (1-\widetilde{M}_{w}) - \frac{\alpha}{\gamma} \ln \frac{\beta -\alpha + \gamma\widetilde{M}_{w}}{\beta -\alpha + \gamma} = \widetilde{t} \, ,  \label{eqn:final_theory}
\end{equation}

where 

\begin{equation}
\widetilde{{M}_{w}} = \frac{M_{w}}{M_{w,i}}, \label{eqn:mw_define}
\end{equation}

\begin{equation}
\widetilde{t} = \frac{t}{\tau}, \label{eqn:t_define}
\end{equation}

\begin{equation}
    \tau = \frac{M_{w,i}}{\pi D_{v,a}Rf(\theta)(c^{0}_{w,s}-c_{w,\infty})}, \label{eqn:tau_define}
\end{equation}

\begin{equation}
    \alpha = (1+A)\frac{c^{0}_{w,s}}{c^{0}_{w,s} - c_{w,\infty}}\frac{M_{g,i}}{M_{w,i}},
\end{equation}

\begin{equation}
    \beta = \frac{M_{g,i}}{M_{w,i}},
\end{equation}

\begin{equation}
    \gamma = \frac{\mu_{g}}{\mu_{w}}\, .
\end{equation}

Here, \(M_{g,i}\) is the initial mass of glycerol in the droplet; \(M_{w,i}\) is the initial mass of water; \(\mu_{g}\) and \(\mu_{w}\) are the molar masses of glycerol and water, respectively; $c_{w,\infty}=c^0_{w,s} \times RH$, where $RH$ is the relative humidity far from the droplet. We used the relative humidity as a fitting parameter between the experiments and the analytical model for particle-laden droplets of water ($w_{g,i}=0$), giving $RH$ = 38 \% in the model (Figure \ref{fig:evaporation_plots}e). This value of $RH$ was used in the simulation and the analytical model for all other values of $w_{g,i}$.

Figure \ref{fig:evaporation_plots}e-i show the excellent agreement between simulations, experiments, and the analytical model. During the initial part of the evaporation, $\widetilde{{M}_{w}}$ decreases almost linearly with $\widetilde{t}$ (Figure \ref{fig:evaporation_plots}). For large times, $\widetilde{{M}_{w}}$ approaches the asymptotic value of $\widetilde{{M}_{w}^*} = \frac{\alpha-\beta}{\gamma}$, when $c_{w,s}-c_{w,\infty}$ becomes zero. The deviations between the simulation and the analytical model can be attributed to the assumption of a perfectly mixed droplet in the analytical approach. Overall, the good agreement of experiments and simulations with the theory shows that, in line with expectations, the evaporation can be modeled as a diffusion-limited process.

\subsection{Formation of the Marangoni Ring}

\begin{figure}
    \centering
    \includegraphics[scale=0.9]{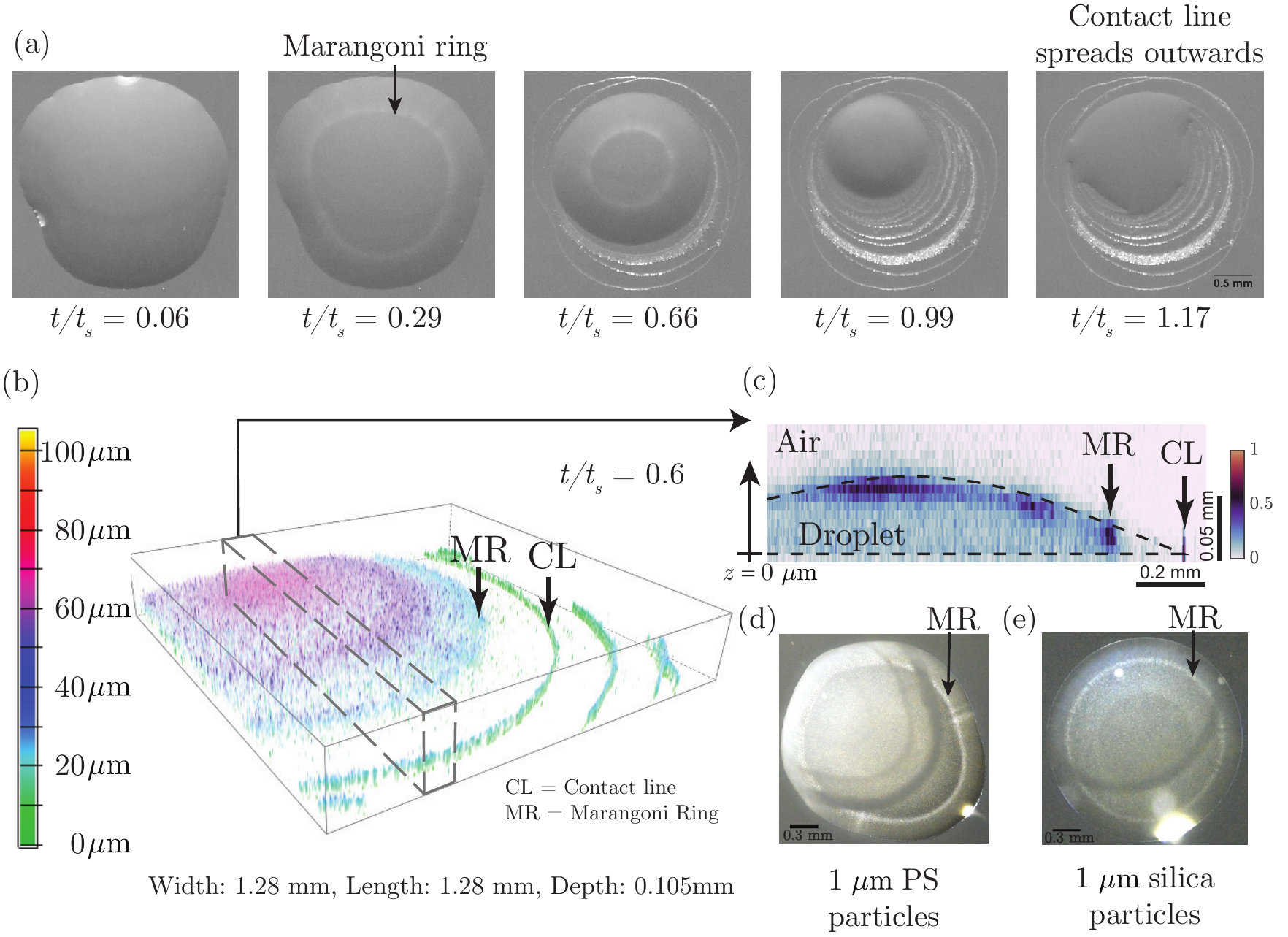}
    \caption{\textbf{Marangoni ring in a particle-laden water-glycerol droplet.} 
    (a) Top-view images of a water-glycerol droplet containing 0.8 \si{\micro \metre} sized silica particles (initial glycerol weight fraction, $w_{g,i}$=5\%). During evaporation, particles accumulate as a ring, termed as Marangoni ring, in between the contact line and the center of the droplet. (b) High-resolution fluorescence confocal microscopy image confirming the Marangoni ring ($w_{g,i}=5 \%$, $t/t_s = 0.6$). The color shows the height of the particles from the substrate. (c) A vertical cross-section of the 3-D image in (b). The color shows normalized fluorescent intensity, which is representative of the particle concentration at a given location. The expected location of the air-liquid interface and liquid-solid interface is marked as a black dotted line. Marangoni ring forms close to the air-liquid interface. (d, e) Top-view images of the Marangoni ring in particle-laden water-glycerol droplets containing (d) commercial 1 \si{\micro \metre} silica particles and (e) commercial 1 \si{\micro \metre} PS particles ($t/t_s=0.4$ and $w_{g,i}=5 \%$).}
    \label{fig:intro_Figure}
\end{figure}

Figure \ref{fig:intro_Figure} shows the formation of the Marangoni ring observed using different experimental methods (for $w_{g,i}=5\%$). During the evaporation of a particle-laden water-glycerol droplet, a bright ring forms in between the contact line and the center of the droplet which we identify as the Marangoni ring (Figure \ref{fig:intro_Figure}a, Supporting Information, Video V1). This ring appears 17 $\pm$ 6 \si{\second} after the drop is deposited on the substrate.  
 
Confocal microscopy shows that this ring corresponds to a dense accumulation of  silica particles (Figure \ref{fig:intro_Figure}b, c, Supporting Information, Video V2). We additionally confirmed that the Marangoni ring is not limited to a particular choice of particles and initial glycerol weight fractions. Both commercial PS particles and commercial silica particles of diameters 1 \si{\micro \metre} form similar Marangoni rings (Figure \ref{fig:intro_Figure}d, e, Supporting Information, Videos V3 and V4). For all these three cases (Figure \ref{fig:intro_Figure} a, d, and e), the initial glycerol weight fraction ($w_{g,i}$) is 5\%. A Marangoni ring also forms for silica particles of diameters 0.8 microns at other initial glycerol weight fractions, \textit{viz.} $w_{g,i}$ = 0.5\%, 25\% and 50\% (Figure S4). The Marangoni ring does not appear during the evaporation of a particle-laden droplet containing only water ($w_{g,i}$ = 0, Figure S5).

Why do particles form the Marangoni ring in a binary droplet? To answer this question, we first look at the flow-field in the droplet.

\subsection{Flow field in the droplet}

\begin{figure}
    \centering
    \includegraphics[scale=0.90]{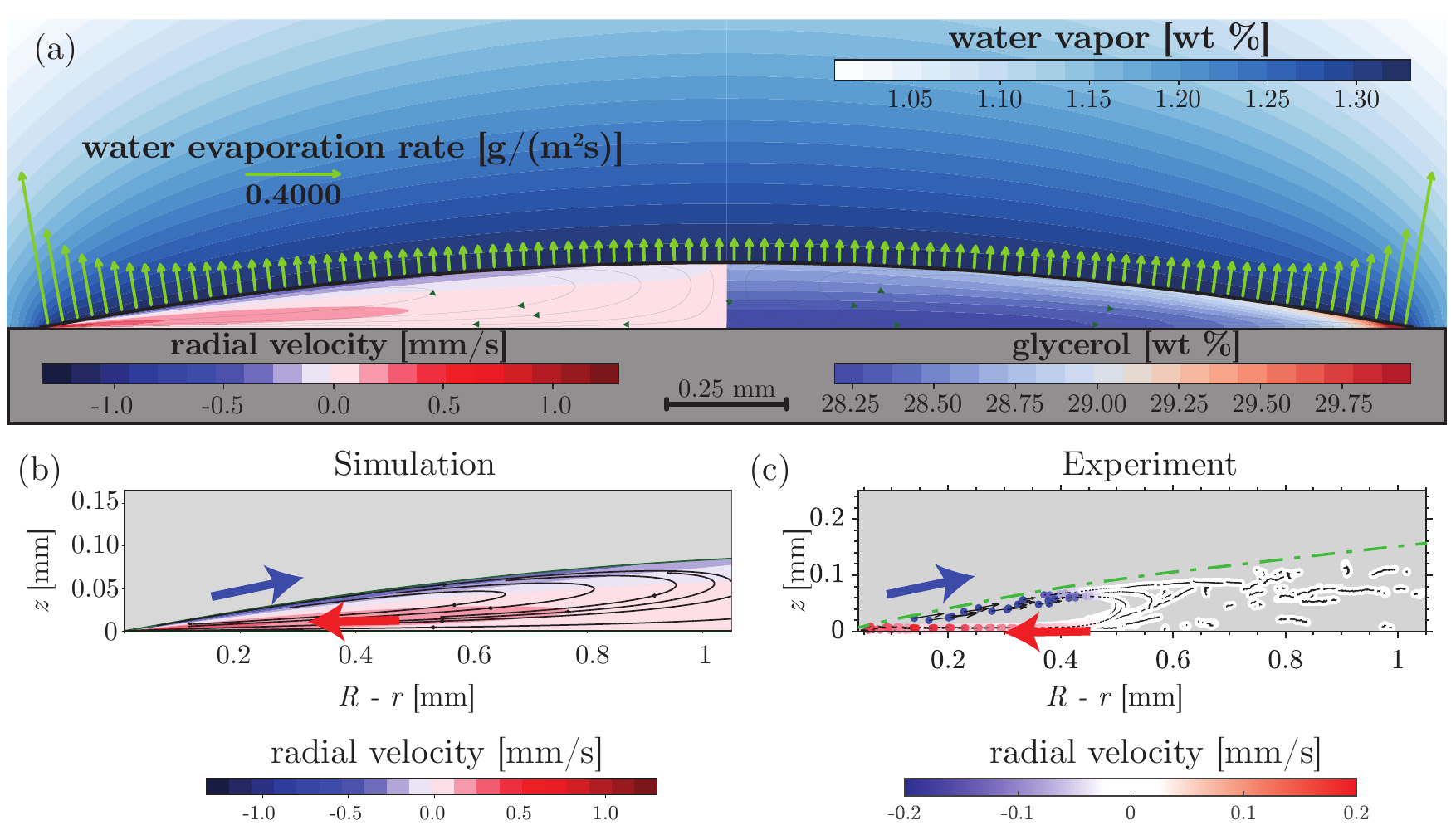}
    \caption{ \textbf{Velocity field in the droplet}. (a and b) Simulations and (c) 3D-PTV measurements ($w_{g,i}=25 \%$) showing the vortex-like three-dimensional velocity field in the droplet. (b) Zoomed-in region of (a) close to the contact-line.
    The flow is radially outward in the lower region of the droplet and radially inwards in the vicinity of the air-liquid interface due to the presence of Marangoni shear. $R$ is the radius of the droplet and $r$ is the radial location from the center of the droplet; thus, $R-r$ shows the distance from the contact line. The velocities are low away from the contact line  ($R-r> 0.5$ mm). For 3D-PTV, a very low concentration ($\mathrm{\approx 5 \times 10^{-5} \%}$ w/v) of fluorescent PS particles was used (see Experiments and Methods). The velocity field in (c) is constructed by superimposing the particle-trajectories over 100 s. The green dash-dotted line in (c) shows the expected location of the air-liquid interface.
    }
    \label{fig:3D_vectors}
\end{figure}

\begin{figure}
    \centering
    \includegraphics[scale=0.9]{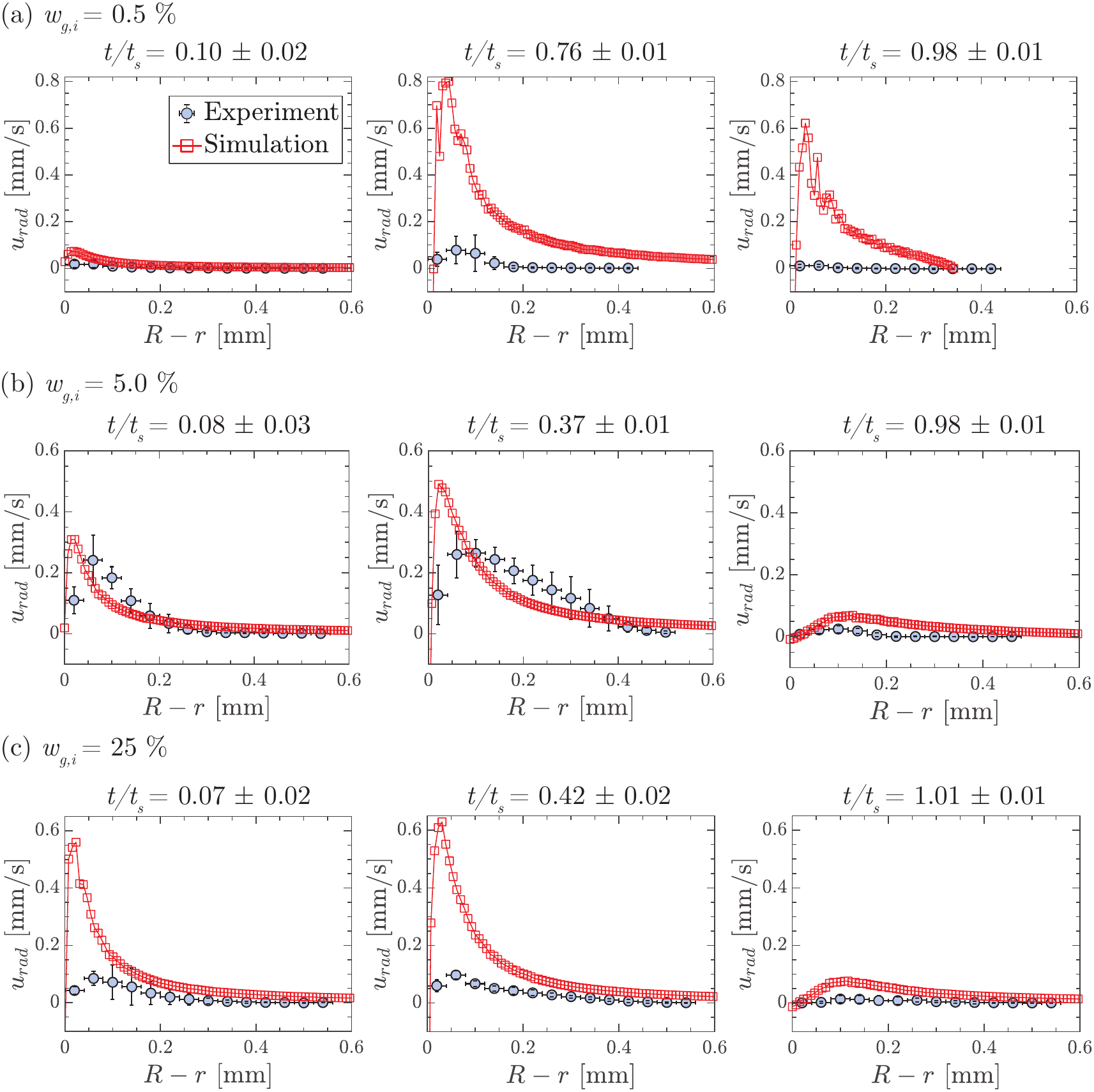}
    \caption{
    \textbf{Radial velocities close to the substrate.} Velocity in a horizontal plane close to the substrate in experiments and simulations, at various time instances, are plotted for a droplet with initial glycerol weight fractions, $w_{g,i}$, of (a) 0.5, (b) 5, and (c) 25\%. $R-r$ is the distance from the contact line. In simulations, for the cases where the air-liquid interface close to the contact-line is deformed strongly by Marangoni contraction (see Supporting Information, Figure S8), $R-r$ refers to the distance from the apparent contact line, instead of the imposed contact line. The experimental results are from \si{\micro PIV} measurements. The velocities plotted for each $w_{g,i}$ are averages from three independent experiments, vertical error bar shows the standard deviation, and horizontal error bars represent the size of interrogation window used in performing \si{\micro PIV}. All the droplets contained 0.1 wt\% of 800 nm fluorescent silica particles at the beginning of evaporation.}
    \label{fig:piv_exp_simulation}
\end{figure}

Particle motion within an evaporating droplet is predominantly governed by the fluid flow. In our water-glycerol droplet, the values of the Rayleigh number $Ra = 2 \times 10^5$, the Marangoni number $Ma = 4 \times 10^5$, and the contact angle $\theta=10^{\circ}$ clearly indicate a surface tension driven Marangoni flow, instead of a buoyancy driven flow (cf. Fig 6a of Diddens et al.\cite{diddens_li_lohse_2021} and the Supporting Information, Section S7 here). This Marangoni flow arises because of differences in volatility and surface tension of water and glycerol, with glycerol being non-volatile and with lower surface tension than water (63 vs. 72 mN/m respectively) \cite{Takamura2012}. Due to the depletion of water at the contact line, a surface tension gradient arises toward the droplet's apex, which leads to a Marangoni shear in the same direction. 
Consequently, the interfacial flow is directed from the contact line toward the droplet apex, as seen in the experiments and simulations (see Figure \ref{fig:3D_vectors}).

The strong interfacial flow, combined with the capillary flow \cite{Deegan1997}, generates a vortical flow structure which we will refer to as ``Marangoni vortex'', typically present in evaporating droplets with strong interfacial flows, regardless of the origin of the Marangoni shear\cite{Hu2005,Still2012,Sempels2013a,Seo2017,Marin2019}. However, simulations show that the Marangoni vortex spans along the whole droplet volume from the contact line to the droplet's center (Figure \ref{fig:3D_vectors}a, Supporting Information, Video V5), while experiments show that the vortex is present mainly close to the contact line (Figure \ref{fig:3D_vectors}b and Supporting Information, Video V6) - in a similar way as observed in other droplets experiencing interfacial Marangoni flows \cite{Still2012,Sempels2013a,Seo2017}.

Figure \ref{fig:piv_exp_simulation} shows a quantitative comparison of the velocity between experiments and simulations, in a plane approximately 10 \si{\micro \metre} above the substrate. 
As we move toward the center of the droplet in Figure \ref{fig:piv_exp_simulation}, the outward radial velocity quickly peaks in the vicinity of the contact line and then decreases at lower rate -- achieving negligible values before reaching the droplet's center. The simulations show a similar trend. However, they yield higher velocity values compared to experiments.
Such a disagreement has been often found in water-based evaporating droplets where strong Marangoni flows are expected due to thermal \cite{Hu2006,Rossi2019} or solutal gradients \cite{Marin2016,Marin2019,vanGaalen2022contamination} and can be explained by the presence of unavoidable interfacial contamination, which reduces the interfacial shear. The presence of interfacial contamination also explains the smaller Marangoni vortex in experiments compared to that observed in simulations. This is discussed in more detail along with additional simulations in the Supporting Information (Section S10).
Nonetheless, despite the quantitative differences, we can conclude that there is a fair overall qualitative agreement between flow-field obtained in experiments and simulations (Figures \ref{fig:3D_vectors} and \ref{fig:piv_exp_simulation}).

\begin{figure}
    \centering
    \includegraphics[scale=0.9]{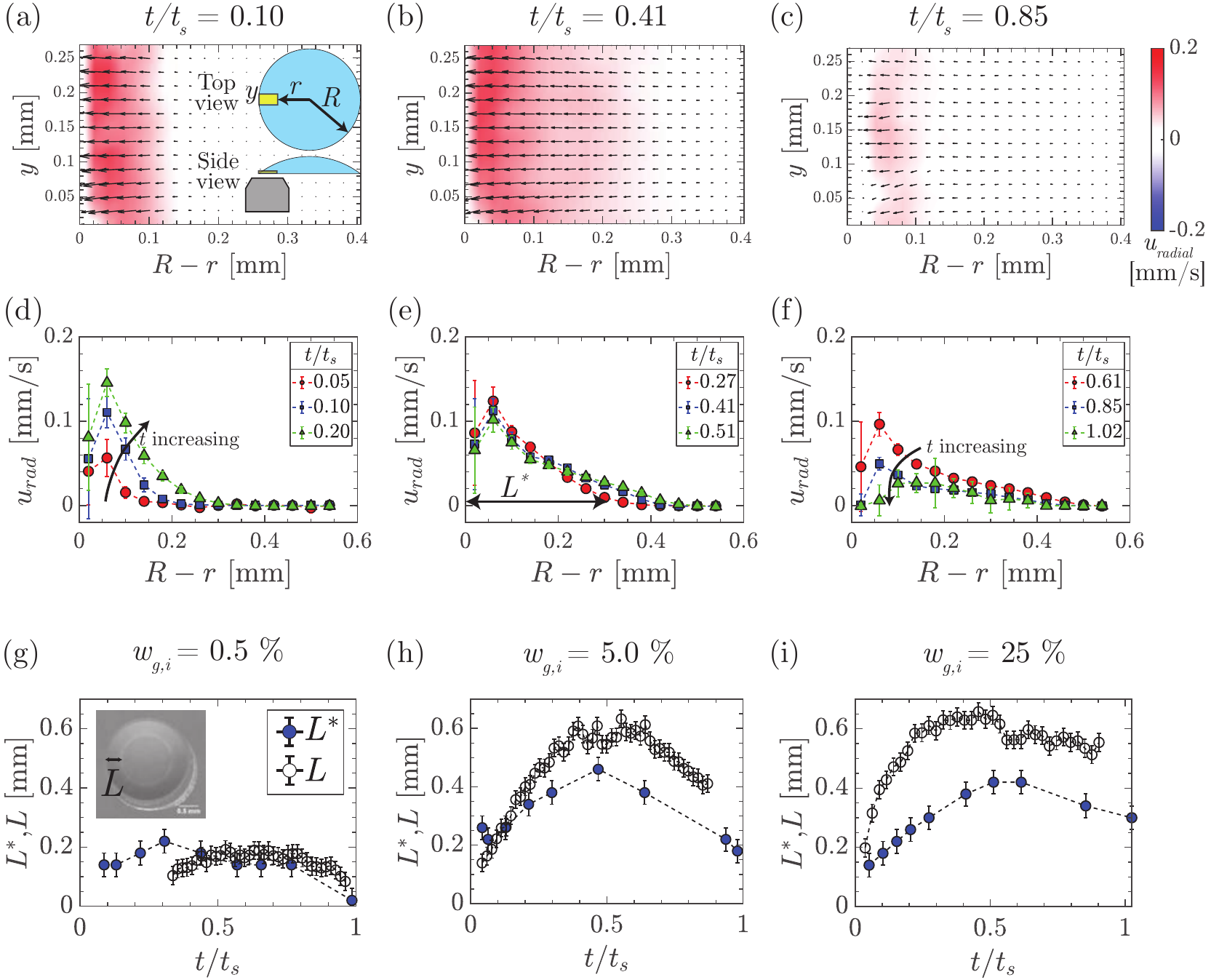}
    \caption{ \textbf{Correlation between Marangoni vortex and Marangoni ring}. (a-c) PIV measurements showing the radially outward velocity ($u_{rad}$) in the droplet near the contact line in a horizontal plane which is approximately 10 \si{\micro \metre} above the substrate, at three different time instants ($t/t_s=0.10,0.41,$ and $0.85$; $w_{g,i}=25\%$). $R-r$ is the distance from the contact line; $y$ axis is along the contact line. (d-f) Plots of the outward radial velocity vs the distance from contact line ($R-r$) at different normalized times ($t/t_s$). $L^*$ is the distance from the contact line where $u_{rad}$ drops below 0.01 mm/s. (g-i) Comparing $L$, the distance of the Marangoni ring from the contact line and the size of the Marangoni vortex $L^{*}$, for a typical droplet with $w_{g,i} = $ (a) 0.5, (b) 5, and (c) 25\%. Additional data and details of error bars in Supporting Information,. $L$ is measured from top-view imaging while $L^*$ is measured from \si{\micro PIV} measurements. Both $L$ and $L^*$ increase with time initially (growth phase) and decrease during the late time (decay phase). All the droplets contained 0.1 wt\% of 800 nm fluorescent silica particles at the beginning of evaporation.}
    \label{fig:piv_2Dvectors_Lstar}
\end{figure}

In order to quantify the temporal changes in the velocity field during the evaporation process, we measure the size of the Marangoni vortex during a droplet lifetime (Figure \ref{fig:piv_2Dvectors_Lstar}a-c) by locating the point where the outward radial velocity drops below a threshold (chosen as 0.01 mm/s, see Figure \ref{fig:piv_2Dvectors_Lstar}e). The distance of this point from the contact line is denoted by $L^{*}$. Figure \ref{fig:piv_2Dvectors_Lstar} shows that the size of the Marangoni vortex initially increases with time and then decreases close to the end of evaporation.

\subsection{Marangoni Ring - Spatio-Temporal Behavior}

Interestingly, the position of the Marangoni ring $L$ from the contact line correlates well in time with the size of the Maragoni vortex $L^{*}$ (Figure \ref{fig:piv_2Dvectors_Lstar}g-i, see Figures S6 and S7 for additional data).
Figure \ref{fig:piv_2Dvectors_Lstar}g-i shows that both $L$ and $L^*$ initially increase with time (growth phase), then reach a maximum, and finally decrease as time approaches the spreading time $t_{s}$ (decay-phase). 
Thus, the location of the Marangoni ring correlates with the radial location where the particle velocity becomes very low. This correlation strongly suggests that the Marangoni ring formation may be governed by the flow-field in the droplet.

\begin{figure}
    \centering
    \includegraphics[scale=0.98]{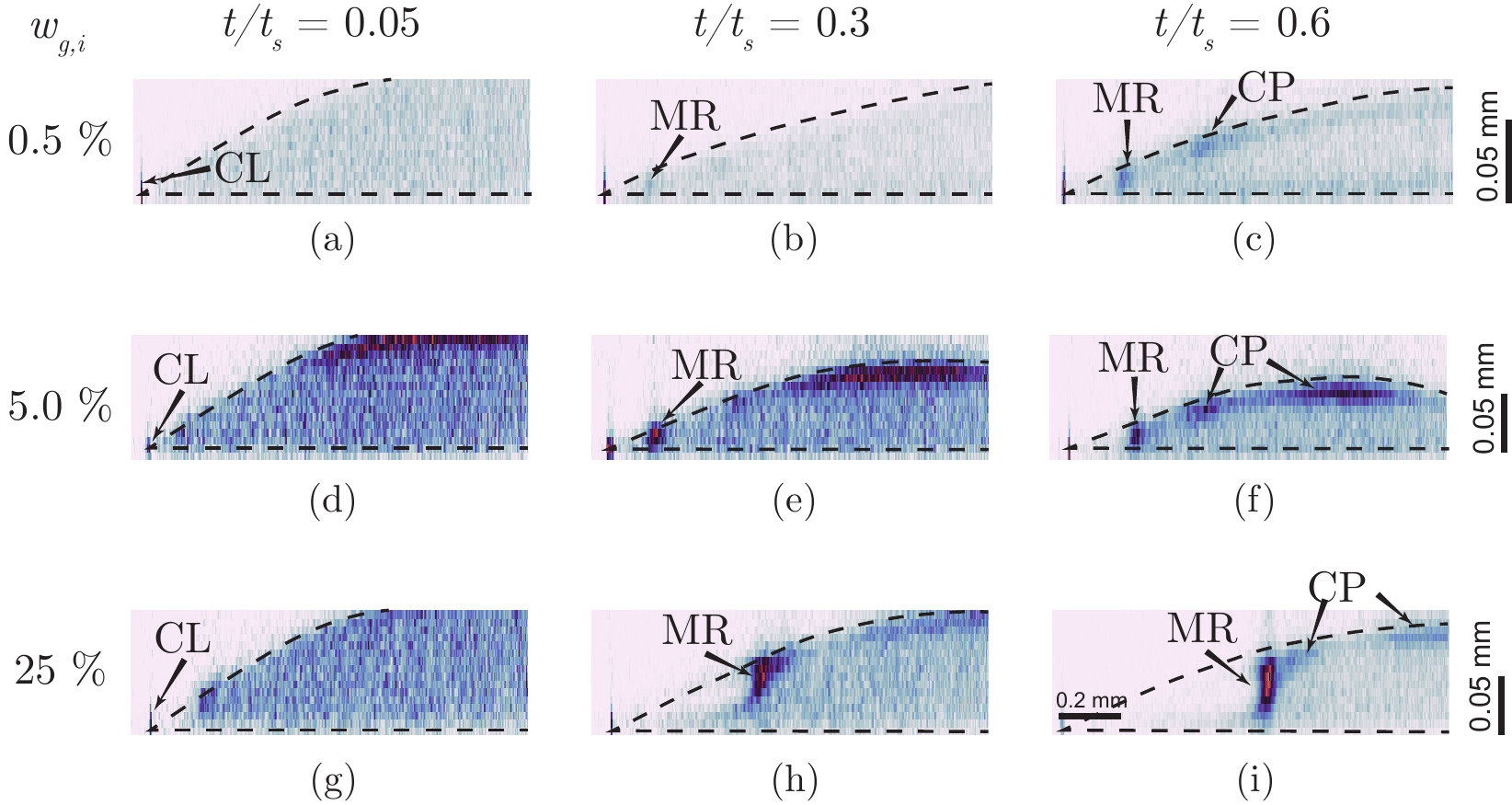}
    \caption{\textbf{Spatial distribution of particles}.
    Normalized fluorescent intensity of 0.8 \si{\micro \metre} silica particles in a vertical cross-section of the droplet observed using fluorescence confocal microscopy, for initial glycerol weight fractions $w_{g,i}=$ 0.5\% (a-c), 5\% (d-f), and 25\% (g-i), at $t/t_s$ = 0.05 (first column), 0.3 (second column), and 0.6 (third column). Darker color indicates a higher particle concentration. The expected location of the air-liquid interface is marked as a black dotted line. However, we note that the precise location of the air-liquid interface cannot be determined with the shown data. The contact line (CL), Marangoni ring (MR), and cap region (CP) are marked in the images. Vertical scale bar: 0.05 mm and horizontal scale bar: 0.2 mm. All the droplets contained 0.1 wt\% of 800 nm fluorescent silica particles at the beginning of evaporation.
    }
    \label{fig:confocal_2d}
\end{figure}

To get more information on the spatio-temporal distribution of the particles inside the droplet, we look at the three-dimensional particle distribution obtained from confocal microscopy of droplets containing the 0.8 \si{\micro \metre} fluorescent silica particles (Figure \ref{fig:confocal_2d}). The particle distribution is nearly uniform in the early moments of evaporation (Figure \ref{fig:confocal_2d}, $t/t_s =0.05$) and eventually, as evaporation progresses, the particles accumulate at different locations along the liquid-air interface (Figure \ref{fig:confocal_2d}). The particle concentration peaks: (1) at the contact line (CL), (2) at the Marangoni ring (MR), and (3) in the vicinity of the droplet's apex, which we call the cap (CP).

Thus, the particles also accumulate to form the cap region between the Marangoni ring and the droplet apex (Figure \ref{fig:confocal_2d}), but usually with lower concentration compared to the Marangoni ring (see Figure \ref{fig:confocal_2d} i, Supprting Information Videos V1, V3, and V4). Because the interfacial velocities are already very low in these regions, we suggest that the cap region is formed mainly because the particles are simply swept by the downward moving liquid-air interface \cite{Bigioni2006,Sperling2014,LiSciRep2016}. A Peclet number of $Pe \approx 50$ in experiments supports the idea that the speed of the liquid-air interface is higher than the diffusion of the particles away from the interface ($Pe = v_{interface}h/D$, where $v_{interface}$ is the velocity of the interface, $h$ is the height of the droplet, and $D$ is the diffusion coefficient of the particles).

\begin{figure}
    \centering
    \includegraphics[scale=0.9]{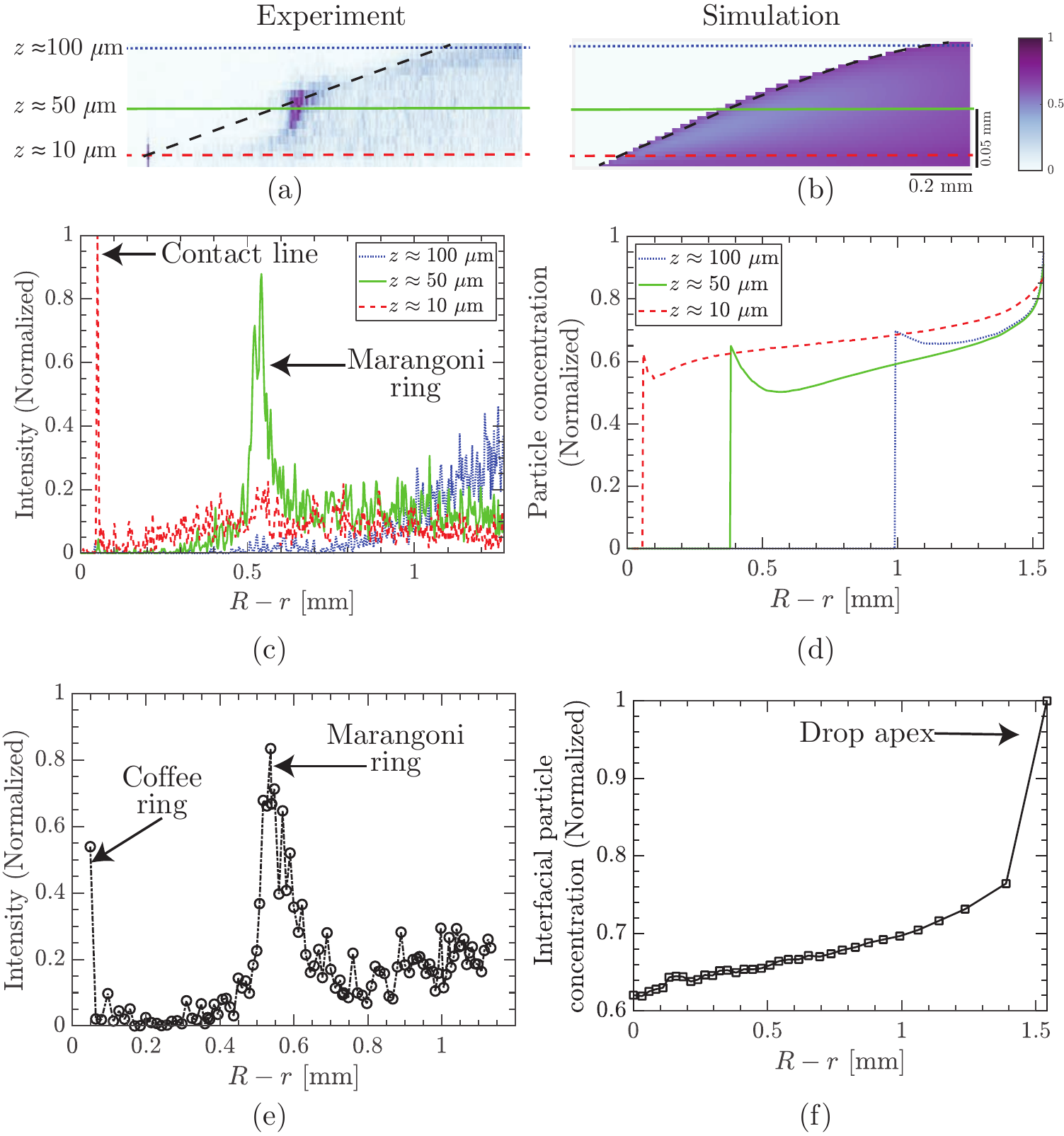}
    \caption{
    \textbf{Particle distribution in a droplet: experiments vs simulation.} (a) Particle distribution inside a droplet visualized using fluorescence confocal microscopy. The colors show the normalized intensity of the fluorescence signal, as indicated by the color bar, and are representative of the concentration of particles. The black dashed line represents the estimated air-liquid interface. (b) Particle distribution inside a droplet as obtained using simulation. The colors show the normalized particle concentration. The horizontal lines in (a, b) show the location of the three planes located at $z=$ 10  \si{\micro \metre}, 50 \si{\micro \metre}, and 100 \si{\micro \metre}. The particle distribution along these three lines are plotted against the distance from the contact line ($R-r$) in (c) and (d). The particle distribution close to the air-liquid interface is plotted for (e) experiments and (f) simulations. The simulations show that the particle concentration increases gradually as one moves from the contact line toward the apex of the droplet. In contrast, experiments show a distinct particle accumulation around 0.5 mm from the contact line, which corresponds to the Marangoni ring. The figure corresponds to a droplet with initial glycerol weight fraction $w_{g,i} = 25\%$ at time $t/ts = 0.3$. 
    }
    \label{fig:particle_distribution}
\end{figure}

To verify whether there is a clear hydrodynamic mechanism behind the particle accumulation, we introduced the particle concentration in the FEM simulation as a continuous field advected by the evaporation-driven flow (see Experiments and Methods).
Figure \ref{fig:particle_distribution} a and b compares the spatial distribution of particles inside the water-glycerol droplet, obtained in experiments and simulations. The experiments show a high particle concentration at approximately 0.5 mm from the contact line and 50 \si{\micro \metre} above the substrate (i.e. the position of the Marangoni ring, Figure \ref{fig:particle_distribution} a, c, and e). However, while FEM simulations show that the particle concentration is high close to the air-liquid interface (Figure \ref{fig:particle_distribution} d), there is no local maxima of interfacial particle concentration between the contact line and the drop apex (compare Figure \ref{fig:particle_distribution} f with e). In fact, simulations predict that the particle concentration at the interface will be highest at the drop apex (Figure \ref{fig:particle_distribution} f). We conclude that the flow field obtained in the simulations cannot reproduce the Marangoni ring formation as seen in the experiments.

There are two main differences between the FEM simulations and the experiments, namely (i) the crucial differences in the flow field and (ii) the lack of non-hydrodynamic particle interactions in the simulations.
Firstly, the Marangoni vortex in experiments is limited to a small region close to the contact line instead of spanning the whole droplet (Figure \ref{fig:3D_vectors}). 
Consequently, instead of being advected by the Marangoni vortex all the way to the drop apex (as seen in simulations), the particles in the experiments are left at some point between the CL and the apex to form the Marangoni ring. 
The most likely reason for the mismatch between the velocity fields seen in the experiments and the simulations is the presence of contaminants in water-based systems, as often invoked and discussed  \cite{Hu2006,Manikantan2020, Kosuke2022,Marin2019}. Impurities can easily neutralize the surface tension gradient, reducing the strength of the Marangoni flow with a small amount of contamination \cite{vanGaalen2022contamination}. Our additional simulations with insoluble surfactants further support this hypothesis (see Supporting Information, Section S10, Videos V5 and V7 for more details).

\begin{figure}
    \centering
    \includegraphics[scale=0.9]{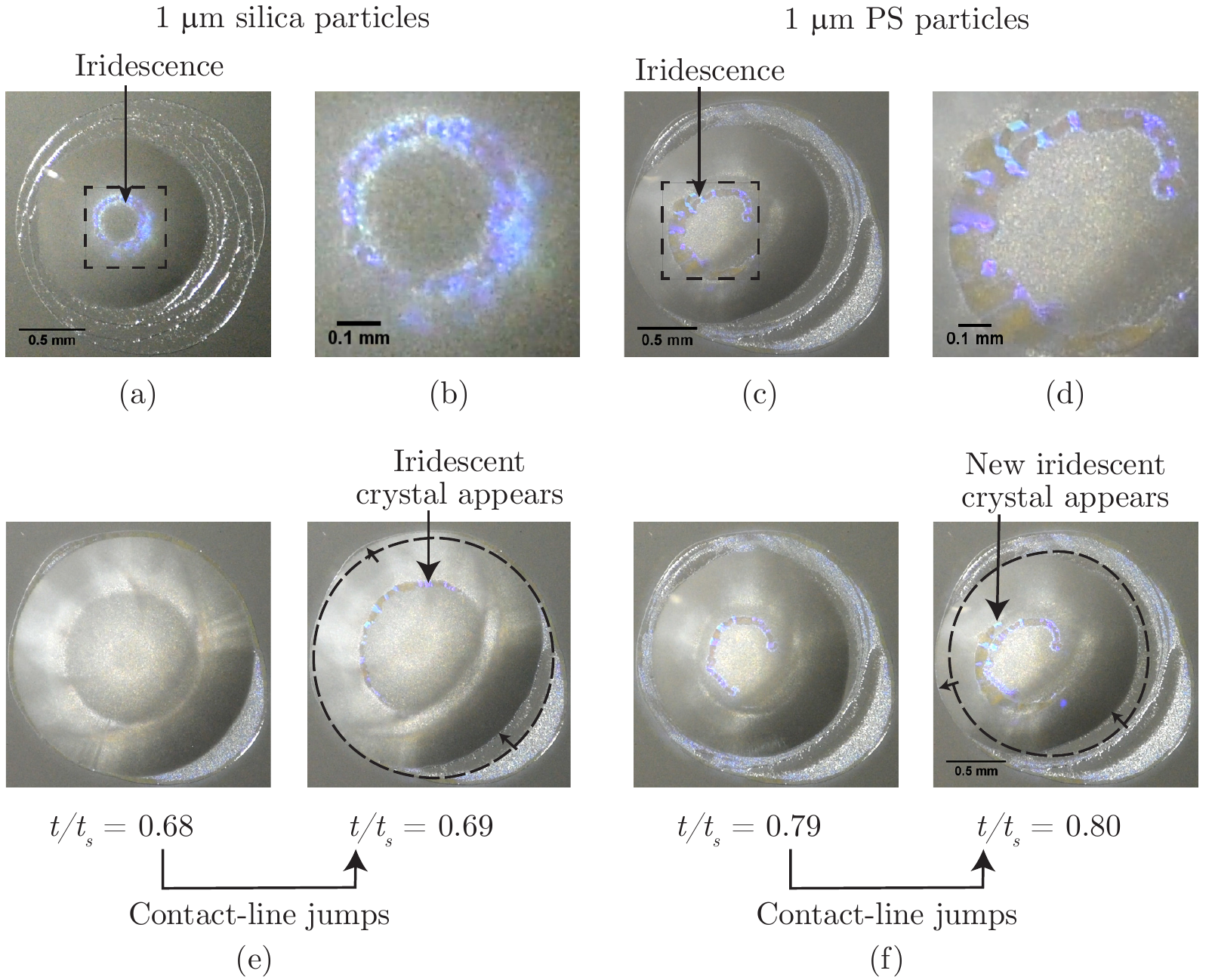}
    \caption{\textbf{Colloidal crystals and iridescence.}
    Experiments with commercial 1 \si{\micro \metre} (a,b) silica particles and (c,d) PS particles, dispersed in water-glycerol droplet ($w_{g,i}=5$\%, $t/t_s = 0.8$) show colloidal crystallization, evident by the iridescence. (b, d) Zoomed-in view of (a, c), respectively. (e, f) Strong correlation between jumps in contact line and the appearance of iridescent crystals for water-glycerol droplets containing the commercial 1 \si{\micro \metre} PS particles. Scale bars 0.5 mm for (a, c, e, f). Scale bar 0.1 mm for (b, d).
    }
    \label{fig:irridescence}
\end{figure}

Secondly, the simulations only include hydrodynamic interactions. However, water-glycerol droplets containing 1-\si{\micro \metre}-diameter silica and PS particles display a remarkable iridescence (Figure \ref{fig:irridescence} and Supporting Information, Videos V3 and V4). The formation of such colloidal crystals is a signature of the closely-packed arrangement of the colloidal particles. The photonic band-gap created by the crystal structure causes the iridescence \cite{kim2011self,Li2016, Hiltner1969}. Interestingly, for PS particles, the formation of these crystals coincides with the jumps in contact line of the droplet (Figure \ref{fig:irridescence}e, f; Supporting Information, Video V3). The formation of iridescent colloidal crystals in the experiments indicates that hydrodynamics is not sufficient to explain the phenomenon.

Future studies should look into controlling the surface properties of the particles, to further understand the colloidal interactions responsible for the Marangoni ring formation. Including colloidal interactions along with hydrodynamics in simulations can also help us to further understand the process \cite{Ren2020,AlMilaji2019,Nguyen2017,Bigioni2006}.

\section{Conclusions}
In this work, we discovered the formation of the Marangoni ring in an evaporating particle-laden water-glycerol droplet. The differences in volatility and surface tension of water and glycerol cause solutal Marangoni flow directed from the contact line toward the droplet apex. However, the interfacial flow loses strength before reaching the apex, and particles are left in an intermediate region, where their concentration increases. This is confirmed by the strong correlation shown between the thickness ($L^*$) of the Marangoni vortex and the position ($L$) of the Marangoni ring from the contact line. Simulations show that hydrodynamic interactions are not sufficient to explain why the local particle density increases to such an extent that colloidal crystals are formed near the liquid-air interface, as can be seen from the iridescence. It is not clear to us what the mechanism is that confines the colloidal particles in our system. Nonetheless, our results provide new insight into particle transport in colloidal droplets which might have important applications such as in diagnostics, inkjet printing, and production of functional coatings, novel opto-electronics, and pharmaceutical products.
 
\begin{acknowledgement}
The authors thank Olga Koshkina and Katharina Landfester for providing fluorescent silica particles. L.T.R. is thankful to Uddalok Sen and Minkush Kansal for fruitful discussions. This work was supported by an Industrial Partnership Programme of the Netherlands Organisation for Scientific Research (NWO), co-financed by Canon Production Printing Netherlands B.V., University of Twente, and Eindhoven University of Technology. D.L. and A.M. acknowledge the funding from European Research Council with Advanced Grant DDD (No. 740479) and Starting Grant (No. 678573) respectively. X.H.Z. acknowledges the support by the Natural Sciences and Engineering Research Council of Canada (NSERC) and Future Energy Systems (Canada First Research Excellence Fund) and the funding from the Canada Research Chairs program.

\end{acknowledgement}

\begin{suppinfo}
\textbf{Section S1-S10:} Experimental setup; contact line motion; details of \si{\micro PIV}; linearization coefficient in evaporation model; Marangoni ring for various initial compositions; top-view visualization of particle-laden water droplet; calculations of Rayleigh and Marangoni numbers; additional data on location of Marangoni ring; additional data on location of Marangoni vortex; hypothetical influence of contaminants.
\\\textbf{Videos}
\\V1: Top view imaging showing Marangoni ring formation in an evaporating water-glycerol droplet containing 0.8 \si{\micro \metre} silica particles
\\V2: Fluorescence confocal microscopy showing Marangoni ring formation in an evaporating water-glycerol droplet containing 0.8 \si{\micro \metre} silica particles
\\V3: Top view imaging showing Marangoni ring formation and iridescence in an evaporating water-glycerol droplet containing commercial 1 \si{\micro \metre} PS particles
\\V4: Top view imaging showing Marangoni ring formation and iridescence in an evaporating water-glycerol droplet containing commercial 1 \si{\micro \metre} silica particles
\\V5: FEM simulations of an evaporating water-glycerol droplet containing (passive) particles
\\V6: Pathlines of particles tracked using GDPTV
\\V7: FEM simulations of an evaporating water-glycerol droplet containing (passive) particles and insoluble surfactants

\end{suppinfo}

\clearpage

\bibliography{achemso-demo}

\end{document}